\documentclass{article}

\PassOptionsToPackage{numbers,compress,sort&compress}{natbib}
\usepackage[preprint]{neurips_2026}

\usepackage[utf8]{inputenc}
\usepackage[T1]{fontenc}
\usepackage{amsmath}
\usepackage{booktabs}
\usepackage{graphicx}
\usepackage{microtype}
\usepackage[x11names,table]{xcolor}
\usepackage{url}
\usepackage{hyperref}
\hypersetup{
  colorlinks=true,
  linkcolor=blue,
  citecolor=blue,
  urlcolor=blue,
  pdfborder={0 0 0},
}
\usepackage{caption}
\usepackage{float}
\usepackage{subcaption}

\title{%
  \includegraphics[width=0.15\textwidth]{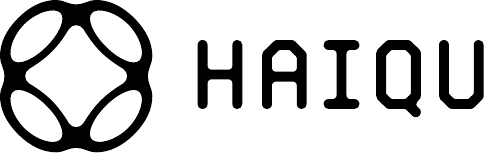}\\[0.5em]
  Demonstration of a Structured Agentic Workflow for Applied Quantum Computing Research
}

\author{%
  {\bfseries Dikshant~Dulal, Maxence~Grandadam, Maciej~Koch-Janusz, Mykola~Maksymenko} \\[0.5em]
  {\normalfont Haiqu, Inc.} \\
  {\normalfont 95 Third Street, San Francisco, CA 94103, USA} \\
  {\normalfont\href{mailto:info@haiqu.ai}{\texttt{info@haiqu.ai}}}
}

\begin{document}

\maketitle
\begin{abstract}
Quantum hardware advances now make it practical to explore quantum approaches to scientific problems. Applied quantum computing research sits at the intersection of domain knowledge, algorithms, theoretical physics, device physics, and quantum and classical computational methods. The breadth of expertise required and the inherent project complexity challenge not only any individual researcher, but also generic AI systems. Here, we introduce the Quantum Agentic Operating System (QAOS), a structured agentic research system designed to address this challenge for quantum and computational R\&D projects. QAOS is built around a nonlinear workflow graph of heterogeneous multi-agent nodes, with a controlled flow of context and computational resources. It decomposes a complex research project into sandboxed multi-agent tasks executed under constraints from the project graph, task-relevant methodologies, and necessary verification steps. We demonstrate QAOS on a quantum chemistry prototyping task, in which it is asked to set up a minimal model for a proton-transfer problem and to implement its solution using sample-based quantum diagonalization (SQD), together with classical reference methods such as FCI (exact diagonalization) and density functional theory (DFT). We compare the QAOS solution to those obtained with baseline agents (Claude Sonnet~5 with the Claude Code harness). In contrast to QAOS, the baseline agents silently ignore explicit instructions, and the runs show large variability.
\end{abstract}

\section{Introduction}

The challenge of executing a sufficiently complex research project often lies not in the difficulty of the atomic tasks, but rather in correctly decomposing the problem into such tasks, as well as ensuring that any scientific and design choices or approximations, performed at the granular levels, remain methodologically consistent throughout and with the original intent. This is particularly pertinent in the context of AI research systems dispatching scores of agents solving individual subtasks.

%A research problem doesn’t inherently specify the procedure needed to compute it. Turning it into executable code requires many decisions that the research question alone leaves open. In computational science, a research brief must be translated into a model of the system, a computational representation, an execution procedure, analytical references, and a method for interpreting the output. Each transformation introduces choices that determine which experiment is actually performed. A resulting computational program may, therefore, execute successfully while answering a different question from the one the researcher intended.

Applied quantum computing research provides a demanding case study of this. Here a typical domain-specific problem, for example in quantum chemistry~\cite{mcardle2020quantumchemistry}, must be formalized and mapped to a model Hamiltonian, which is then solved through a multi-stage procedure combining analytical methods with classical and quantum numerical methods. 
It is essential to ensure that models and approximations chosen do indeed capture the phenomena studied and are feasible to simulate, and the computational methods used are not only correctly implemented but also accurate within the regime studied. Failure to explicitly enforce this \emph{global consistency} in an AI system often results in runnable code, confident analyses, and ``results'', which are, however, scientifically worthless.

%Quantum chemistry provides a demanding case study of this general problem. The geometry and reaction coordinate must first be fixed; a basis and electronic space must then be chosen; the resulting fermionic Hamiltonian must be mapped to qubits; a circuit and sampling procedure must be defined; and the sampled configurations must finally be converted into an energy and an observable. If one of these steps changes silently, two individually reasonable calculations may no longer describe the same Hamiltonian or even the same molecular structure~\cite{mcardle2020quantumchemistry}.

Such silent failures are especially likely when coding agents assemble the pipeline. %A single generation pass can invent a geometry, freeze a core, or alter the mapped Hamiltonian while still returning a runnable program. 
Recent scientific-agent benchmarks show that agents can recover parts of computational workflows, but also that they produce incomplete procedures, methodological deviations, and execution failures~\cite{chen2024scienceagentbench,huang2026automat,duston2025ainsteinbench}. Other work organizes scientific agents as explicit graphs of tasks, state, and provenance~\cite{feng2026graphengineering,huber2020aiida,gottweis2026accelerating}. The closest system, El Agente Gr\'{a}fico, provides a typed execution runtime that validates and transfers scientific state between agents~\cite{bai2026agente}, and related work confines model nondeterminism within validated workflow graphs~\cite{balis2026researchquestion}. However, these systems are domain-specific, and they miss a systematic mechanism for formalizing a loose scientific problem specification in the broad applied quantum domain, decomposing it into independently verifiable claims and baselines, and enforcing them throughout implementation and execution as constraints and tests.

The Quantum Agentic Operating System (QAOS) addresses this gap, unfolding the project into nodes of a heterogeneous graph, with tightly controlled information and computational flow. Initial analysis and formalization nodes turn the brief into a detailed experimental plan, with a formalized research question, model systems and relevant methods, as well as reproducible references. These are locked before implementation and approved by a human researcher. Later stages convert that plan into verification protocols, integration tests, reference analytical calculations and implemented prototype blocks and components, which are carried out by sandboxed subagents, and adjudicated by orchestrators.

%as a graph workflow system for quantum computing research. Instead of asking one agent to carry the whole problem through a single coding session, it organizes the work as scientific stages connected as a complex graph. The early stages turn the brief into an explicit experimental plan: the research question is formalized, relevant methods and reproducible references are gathered, and a short project vision records what will be built and which protocol choices are locked before implementation. Later stages convert that approved plan into verification protocols, integration tests, reference analytical calculations and implemented prototype blocks and components. Human-in-the-loop (HITL) review sits at the transitions that shapes the direction of the experiment. 

Here we demonstrate QAOS on the example of a quantum chemistry prototyping task, where the research brief is to investigate the inclusion of a quantum algorithm, sample-based quantum diagonalization (SQD)~\cite{robledomoreno2025chemistry}, in a larger classical computational pipeline simulating proton transfer, ultimately of interest in battery chemistry. QAOS suggests a minimal model system that is physically meaningful and amenable to simulation with both SQD and multiple classical methods, providing reference benchmarks, and, upon approval of the plan, builds and executes the complete simulation.

%We demonstrate the system on proton transfer in $\mathrm{H_5O_2^+}$, a chemistry use case chosen because a small full-space electronic problem still admits an exact classical referee while remaining relevant to quantum solvers. The rest of the paper makes two points. First, when the quantum research pipeline is split into modules with fixed jobs and verification checks at their interfaces, the silent geometry and electronic-space failures seen in baseline LLM runs do not appear or are prevented by automated verifications. Second, under that modular, test-driven build, the resulting sample-based quantum diagonalization (SQD) prototype recovers a barrier closest to the exact classical answer among the approximate methods compared in this study.

\section{Assembling a testable quantum protocol}\label{sec:protocol}

\begin{figure*}[t]
  \centering
  \includegraphics[width=\textwidth]{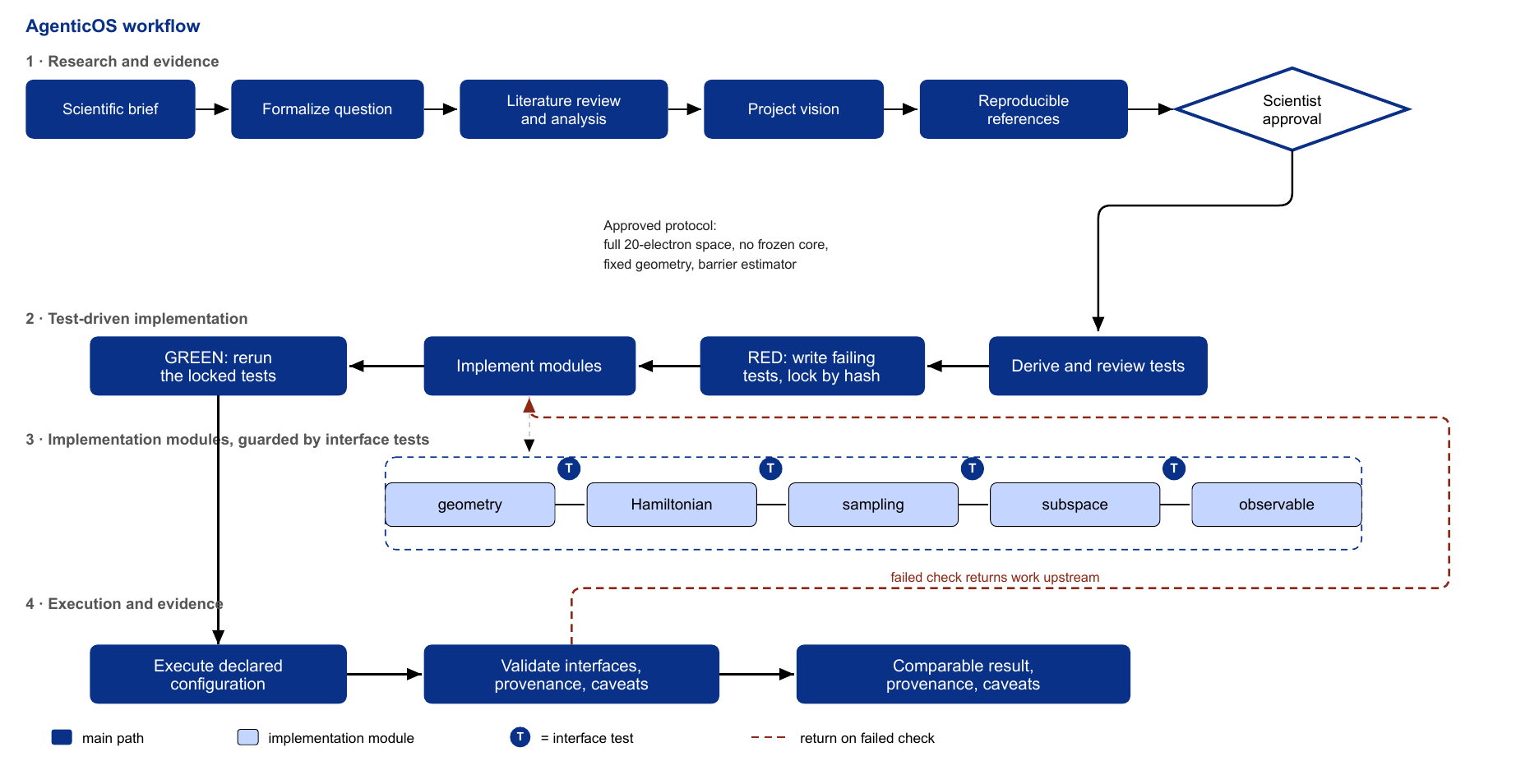}
  \caption{Schematic view of the Quantum Agentic OS workflow used in this demonstration. Research and evidence stages (dark) generate a formalization, define a build protocol, and specify reference results to reproduce, which are human-approved. Test-driven implementation converts the protocol into immutable tests before coding. The implementation nodes (light) are guarded at each interface by the tests. Execution validates the interfaces and the run record, and a failed check returns work to the relevant upstream stage. Node labels and arrows remain legible in grayscale.}
  \label{fig:workflow}
\end{figure*}

QAOS does not attempt to solve the problem in a single agentic session. Instead, it decomposes the project by progressively building a graph of nodes (each a multi-agent module in itself), where each completed node contains structured records serving as a specification and information source for the connected descendants. Inside each such node, an orchestrating agent holds the stage goal and assigns bounded subtasks to specialist agents. One or more critic agents review the drafts before a structured record is written, and at defined transitions a human researcher approves the scientific direction. Downstream nodes receive that record, not the internal agent dialogue. An example of this workflow for the test problem described below is shown in Fig.~\ref{fig:workflow}. The key problem formalization node derives the high-level problem specification, identifying the relevant physical objects (Hamiltonians, observables, etc.), methods, and computations to be performed. It is aided in these tasks by specialized literature review nodes. The project vision node then derives a more detailed description of which components and reference baselines need to be built and verified to achieve the goal, and which protocol decisions are fixed before code generation. This plan, once approved or amended by the human researcher, is then executed by the agents in the downstream nodes. These can be either analytical, involving exact computation, or, as in the test case projects considered here, numerical, with agentic code generation.

%QAOS does not treat the demonstration as a single agentic session. The executed run first builds the scientific project workflow of multi-agent steps that the implementation must obey. Problem formalization states the question and the observable. Literature review and critical analysis collect the methods and evidence that justify the computational choices. A project vision then records, in short form, what prototype will be built, which components and reference baselines need to be built and verified to achieve the goal, and which protocol decisions are fixed before code: here, the proton-transfer toy, the full electronic space, and the barrier as the target quantity. Reproducible-reference module attaches the concrete method specifications that the build for each prototype component must follow. The early stages exist so that geometry, electronic space, and evaluation are chosen and verified deliberately rather than invented during coding. Their connected outputs become the inputs to prototype construction flow after a human approval gate. Figure~\ref{fig:workflow} shows how this research path joins the test-driven build and the quantum implementation path.

QAOS reduces the risk of silent divergence from objectives by using test-driven development~\cite{wilson2014bestpractices,balaban2021tenrules} while coding. Approved protocols and baselines are turned into immutable unit and integration tests before any code is written. The prototype stage first records scientific requirements, constraints, assumptions, and test scenarios in the red development phase. An orchestrating agent dispatches implementing agents against the fixed specification, and verifies compliance with the tests in the green phase.

\section{Quantum-chemistry case study}\label{sec:casestudy}

\paragraph{From the chemical problem to an executable quantum algorithm.}
In this demonstration QAOS was tasked with reproducing a classical quantum chemistry workflow in materials science, replacing the critical step that solves the correlated electronic problem with a suitable quantum algorithm. The QAOS prototype node suggested using the Zundel cation $\mathrm{H_5O_2^+}$ with a fixed inter-oxygen distance as a minimal model of proton transfer in water~\cite{schroder2022hydratedproton}. The computational workflow uses Density Functional Theory (DFT) to compute the equilibrium positions of other atoms, a quantum or classical solver to obtain the electronic ground-state energy, and then calculates the correction to the proton diffusion barrier due to electronic correlations. A minimal basis set for the electrons was used (STO-3G) and the oxygen--oxygen separation was fixed at an arbitrary $d=2.9$~\AA. Details of the molecular model and the finite-basis referee are given in App.~\ref{sec:appC}. The same specification was given to the baseline Claude agents (App.~\ref{sec:appB}).

%directly given to the baseline agents runs such that they do not have to re-discover it and can focus on the implementation. More details on the QAOS workflow and baseline agents prompt can be found in App.\ref{sec:appA}

%This is a controlled, exactly solvable, toy model.

%At the start of the prototyping module, the QAOS settles, after review by the user, on a study of the Zundel cation $\mathrm{H_5O_2^+}$, a minimal model of proton transfer in water~\cite{schroder2022hydratedproton} for a fixed inter-oxygen distance and the full workflow uses multiple specific tools: Density Functional Theory (DFT) for relaxing computing the equilibrium positions of the other atoms, a quantum or classical solver to find the minimum electronic configuration, calculation of the correction induced by the electronic correlation to the diffusion barrier for the proton. We treat this as a controlled, exactly solvable, toy model. We use the minimal basis set to represent the electonic degree of freedom (STO-3G) and the oxygen-oxygen separation is fixed at an arbitrary d=2.9~\AA. The specification of this exact physical system are directly given to the baseline agents runs such that they do not have to re-discover it and can focus on the implementation. More details on the QAOS workflow and baseline agents prompt can be found in App.\ref{sec:appA}

In the chemistry problem studied, for each proton position $\delta$ along the shortest path connecting the two oxygen atoms, a second-quantized molecular electronic Hamiltonian needs to be constructed,
\begin{equation}
  \hat H(\delta)=E_{\mathrm{nuc}}(\delta)
  +\sum_{pq\sigma}h_{pq}(\delta)a^\dagger_{p\sigma}a_{q\sigma}
  +\frac{1}{2}\sum_{pqrs\sigma\tau}g_{pqrs}(\delta)
  a^\dagger_{p\sigma}a^\dagger_{q\tau}a_{s\tau}a_{r\sigma},
  \label{eq:hamiltonian}
\end{equation}
where $E_{\text{nuc}}$ is the classical Coulomb potential from the nuclei, $a_{p\sigma}$ ($a^{\dagger}_{p\sigma}$) is the annihilation (creation) operator for an electron of spin $\sigma$ in the orbital $p$, $h_{pq}$ is the tunneling probability between orbital $p$ and orbital $q$ and $g_{pqrs}$ is the electron--electron repulsion integral between orbitals $p$, $q$, $r$ and $s$.

The Zundel cation is represented in a minimal set of 15 orbitals (STO-3G basis), filled with 20 electrons. Full configuration interaction (FCI), i.e., exact diagonalization of the Hamiltonian in Eq.~\eqref{eq:hamiltonian} at each position $\delta$, gives the exact energy barrier: 574~meV. The complexity of the traditional quantum chemistry workflow is illustrated by the variety of solutions obtained by numerical methods treating the electronic correlations (last term in Eq.\eqref{eq:hamiltonian}) in different approximations. DFT collapses the barrier to 83~meV, missing nearly 500~meV. A mean-field Hartree--Fock approach overshoots at 785~meV, while a coupled-cluster singles and doubles (CCSD) calculation recovers most of the correlation and gives 606~meV.% All comparisons use the same electronic Hamiltonian, but treat the electronic correlations (last term in Eq.\eqref{eq:hamiltonian}) at different degrees of approximation.

\begin{figure*}[t]
  \begin{subfigure}[b]{0.5\textwidth}
    \centering
    \includegraphics[width=0.95\textwidth]{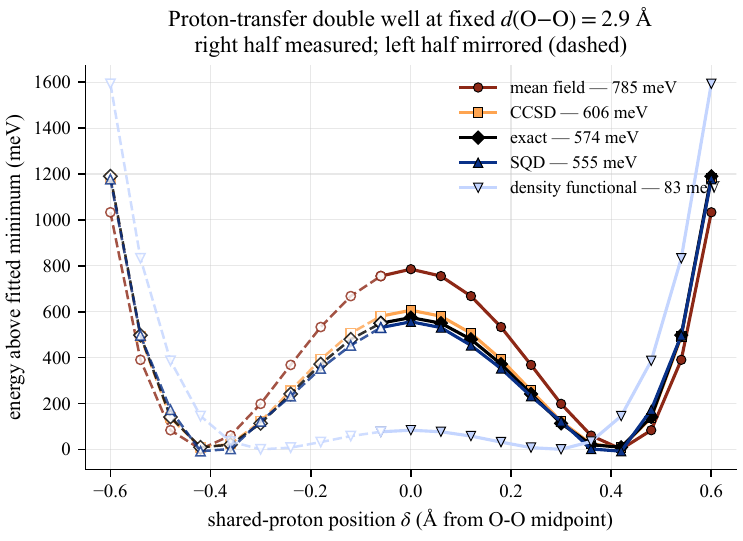}
  \end{subfigure}
  \hfill
  \begin{subfigure}[b]{0.5\textwidth}
    \centering
    \includegraphics[width=0.95\textwidth]{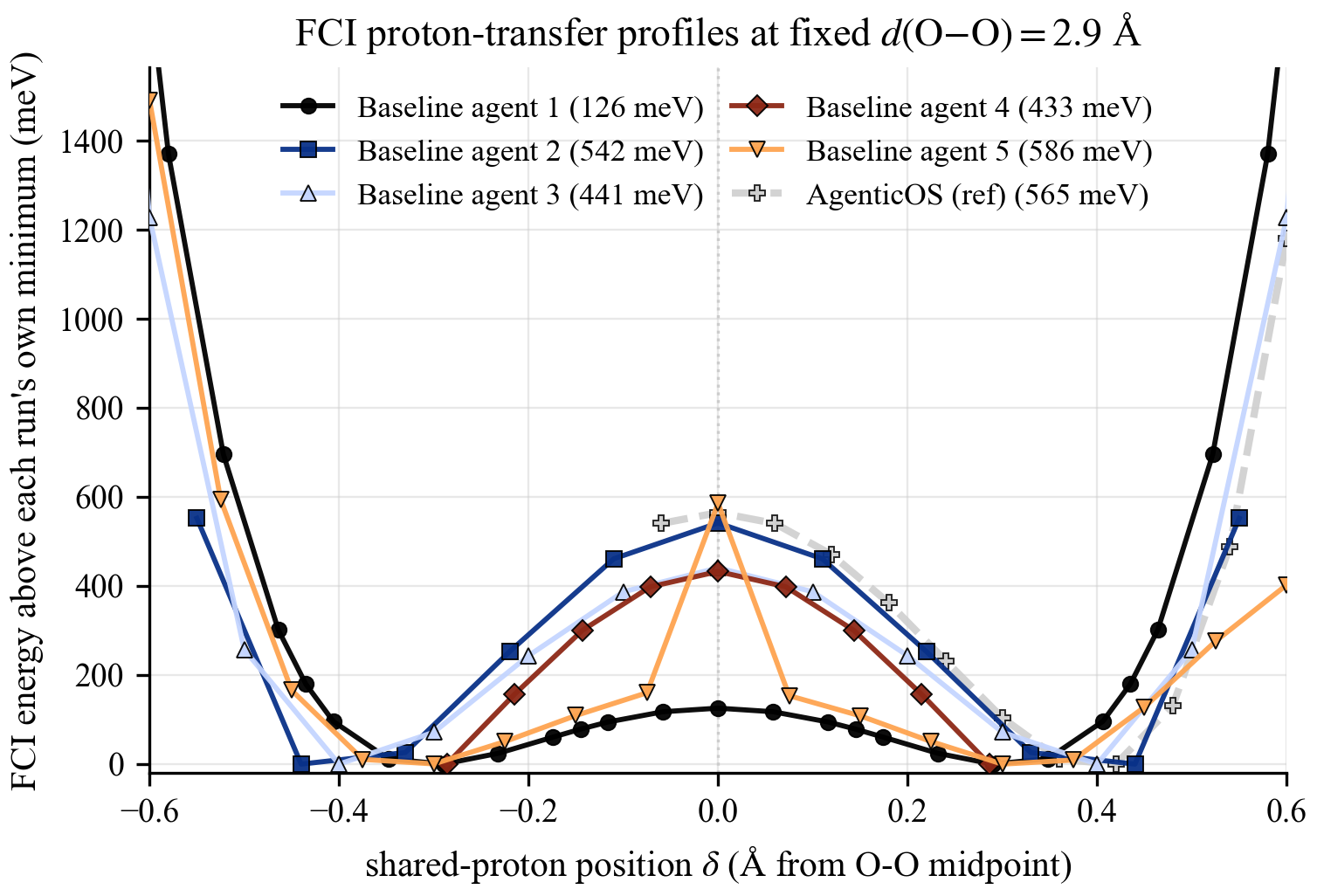}
  \end{subfigure}
  \caption{\textbf{Left:} Potential-energy profiles produced by the QAOS-guided protocol. FCI is exact within the STO-3G space; the quantum SQD method recovers the energy barrier to within 20~meV. \textbf{Right:} Five baseline FCI attempts using the standard Claude Code harness with the same Sonnet~5 model as QAOS. The baseline agents demonstrated large variability, breaking the brief by choosing different electronic subspaces and geometries, committing critical mistakes before even reaching the quantum algorithm stage. Legend labels identify each method for color and grayscale print.}
  \label{fig:potential-curves}
\end{figure*}

%Adding to this a quantum algorithm requires a firm handle of the different classical techniques to pick a correct baseline comparison. 
In our example, we are interested in substituting a classical component with a quantum algorithm, and, following the suggestion of QAOS, we implement the \emph{Sample-based Quantum Diagonalization} (SQD)~\cite{robledomoreno2025chemistry}. It asymptotically converges to the exact FCI result, and is feasible at the required scale of 30 qubits. A classical trial state is prepared, a local unitary cluster Jastrow circuit is sampled $10^6$ times at each geometry, and a Hamiltonian built from those samples is diagonalized to approximate the ground-state energy. The complete workflow built with QAOS gives a 555~meV barrier for SQD, within 20~meV of the exact answer. This is shown in Figure~\ref{fig:potential-curves} (left), where different methods (also implemented by QAOS) are compared. The barrier calculations are in line with the existing literature for this problem~\cite{Dagrada2014}.

\section{Silent failures in baseline LLM runs}\label{sec:baselines}

While this paper is intended as a demonstration only, we also ran a limited baseline comparison of QAOS with a generic unstructured agentic setup, using the same 
Claude Code harness and Claude Sonnet~5 LLM with Medium effort settings. We provided the same supporting paper and baseline prompt given in App.~\ref{sec:appB} explicitly instructing to keep the full electronic space.
Human researchers answered only the questions Claude asked.

Despite these instructions the five baseline attempts selected distinct geometries (\emph{e.g.}~allowing the oxygens to move, or the proton position to veer off the shortest path). Furthermore, in three of the five, the model violated the full-electronic-space instruction and froze core electrons. The right panel of Figure~\ref{fig:potential-curves} shows the five profiles together with the QAOS curve under its locked geometry and electronic space. QAOS keeps those choices from drifting by interviewing the user, obtaining approval of the geometry and electronic space before implementation, and checking those locks in the corresponding nodes with a test at every interface before a barrier is accepted.

\section{Conclusion}
Our main contribution is the Quantum Agentic Operating System (QAOS), a structured agentic workflow for applied quantum computing research in which scientific decisions are set before implementation and subsequently enforced through problem decomposition and interface verification. In the proton-transfer case study, this produced a reproducible computational experiment and an SQD algorithm estimate close to the exact finite-basis result. Rather than relying on a single agent to remain scientifically consistent over a long session, QAOS constrains model non-determinism within a verifiable research process. 

%We use a minimal basis, and the quantum energies come from noiseless sampling of a fixed subspace rather than noisy quantum hardware or the full iterative sample-based procedure. Even so, two results stand out. Splitting the quantum pipeline into modules with fixed jobs, and locking those jobs with tests before implementation, keeps the geometry and electronic-space failures of the baseline Claude runs from passing as a finished experiment. Under that same modular, test-driven build, SQD, the quantum method in the pipeline, recovers a barrier closest to the exact classical answer among the approximate methods shown here. The same mid-tier model, Sonnet~5 Medium, was used inside AgenticOS and in the five baseline Claude runs: the workflow path still produced a locked experiment that can be checked against an exact answer on that faster, lower-cost model. Future work can extend the same protocol discipline to hardware backends and larger active spaces.

\paragraph{LLM use disclosure.}
Claude Sonnet~5 with Medium effort settings and Claude Code harness were used within QAOS for research and implementation stages, and in the baseline agent comparison runs. Human researchers selected and approved the scientific protocol, reviewed the generated artifacts and tests, executed the calculations, checked the reported evidence, wrote and revised the manuscript. In the baseline agent runs, the human answered only the questions Claude asked and did not add further scientific steering. An LLM was also used for preparing and editing the manuscript.

\newpage
{\small
\bibliographystyle{unsrtnat}
\bibliography{references}
}

\appendix
\section{Supplementary material}
\subsection{QAOS workflow graph used in the paper}\label{sec:appA}
\begin{figure}[H]
    \centering
    \includegraphics[width=0.95\linewidth]{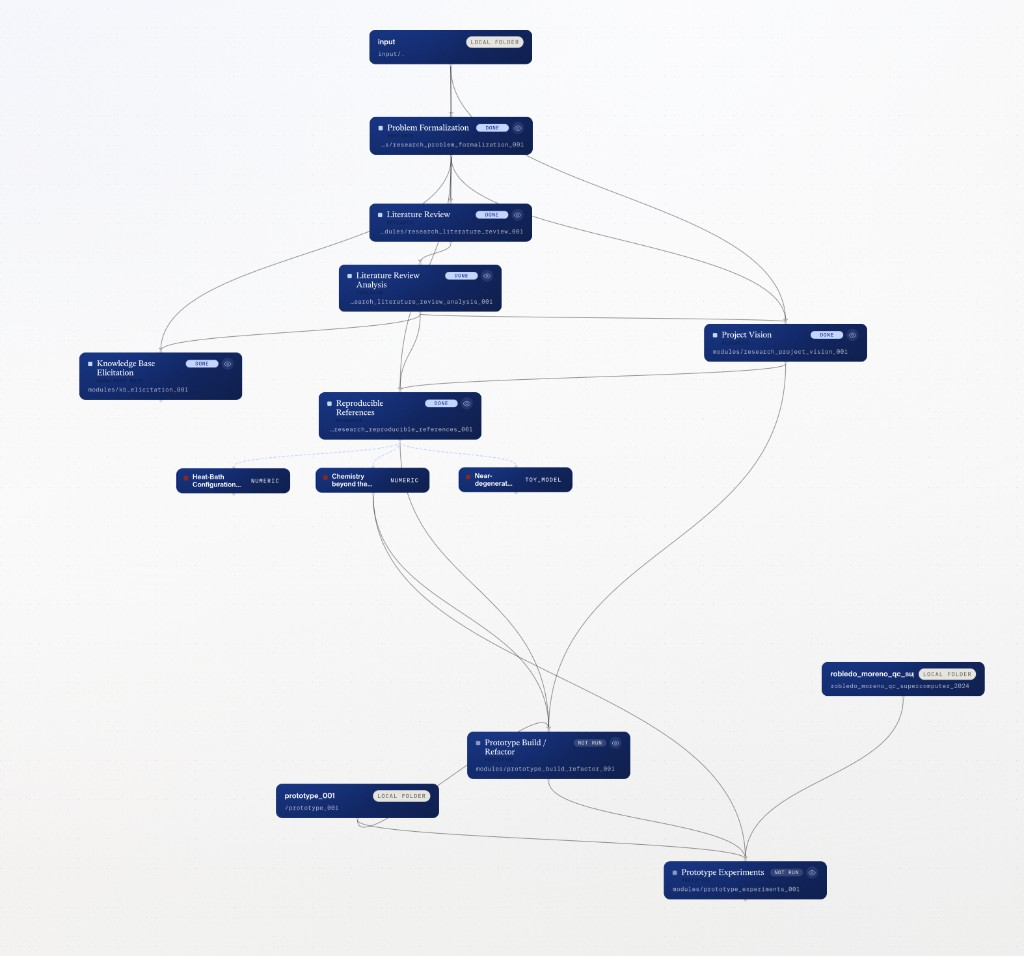}
    \caption{Project graph for the proton-transfer demonstration as shown in the Quantum Agentic OS interface. Each node is a multi-agent module. Edges show the controlled flow of context used to assemble the experiment.}
    \label{fig:screenshot}
\end{figure}

Figure~\ref{fig:screenshot} shows the project graph of nodes for the proton-transfer demonstration, as displayed in the Quantum Agentic OS interface. As in Section~\ref{sec:protocol}, each node is a multi-agent module in itself (or a local input folder attached to the graph). Edges indicate which completed outputs become available as context to later stages. Figure~\ref{fig:workflow} is the schematic of the same organization. This appendix figure is the concrete graph used for the case study.

As defined in Section~\ref{sec:protocol}, each multi-agent node follows the same internal pattern: an orchestrating agent holds the stage goal and dispatches specialist subtasks, critic agents review drafts before a structured record is written, and a human researcher approves direction at defined gates. Downstream nodes receive that record, not the internal agent dialogue. The walkthrough below notes how that pattern appears at each stage of this run.

The run begins from a local \textbf{input} folder containing the project brief and a paper on SQD, the quantum method used in the QAOS run. That material motivates the study and the choice of solver, but it does not yet specify the demonstration molecule, basis, or full pipeline. This folder is not a multi-agent node. It is read-only kickoff material for formalization.

The \textbf{problem formalization} node turns the brief into a written problem statement: the target observable, a candidate computational workflow, main sources of uncertainty, and scope limits. Inside the node, the orchestrator ingests the brief and assigns specialist agents to analyze inputs, check citations, and draft the formalization and notation. Critic agents review claims and epistemic labels before the structured problem statement is written. Downstream nodes treat that record as the working definition of the scientific question (Section~\ref{sec:protocol}).

\textbf{Literature review} nodes compile papers relevant to that scope. Inside each review node, the orchestrator runs scoped search agents over query families tied to the formalization and compiles ranked short memos. Critic agents filter relevance to project scope so the outward product is a literature index, not raw search logs.

The \textbf{literature review analysis} node then judges which claims in that literature are solid enough to rely on, and which steps of the planned computational workflow are still weakly supported. A synthesis specialist turns the memos into a critical review of supported claims, gaps, and untested transfers. Critic agents challenge overreach and missing caveats. That assessment guides planning and the knowledge-base stage.

The \textbf{knowledge-base elicitation} node records methods, observables, and open questions so later stages need not reread long documents (Section~\ref{sec:protocol}). Extractor specialists pull entities and relations from the formalization and analysis. Critic agents drop unsupported links and merge cross-artifact connections, and the researcher accepts the ontology schema before structured knowledge-base pages are written for reuse.

The \textbf{project vision} node combines the formalization and literature synthesis. Inside the node, a vision specialist proposes candidate directions while the orchestrator runs a short interview with the researcher. Decisions on project direction, success criteria, and parked alternatives are frozen in the vision record before build. This is the human approval step before implementation emphasized in Sections~\ref{sec:protocol} and~\ref{sec:baselines}. Vision fixes direction. It does not yet select the minimal molecular model used in the figures.

The \textbf{reproducible references} node selects a small set of published methods that the later build must be able to reproduce or justify. In this project these included classical high-accuracy electronic-structure solvers and sample-based quantum diagonalization (SQD), the quantum method used in the QAOS runs of Section~\ref{sec:casestudy}. Reference-scout specialists search and draft reference cards. The orchestrator verifies doubtful metadata before critics and a researcher selection gate approve a minimal set. On the interface graph, each chosen method appears as its own attached reference node under reproducible references. Those attachments are selected cards, not separate multi-agent nodes.

The \textbf{prototype build~/~refactor} node then uses the approved vision, the selected references, and local prototype code. At this stage the demonstration protocol is fixed for coding: Zundel $\mathrm{H_5O_2^+}$, fixed oxygen-oxygen separation, STO-3G with the full electronic space, classical solvers plus noiseless SQD, and the proton-transfer barrier as the primary observable (Section~\ref{sec:casestudy}). Inside the node, the orchestrator turns the approved protocol into requirements, assumptions, and immutable tests in the red phase, then dispatches implementing agents against that specification in the green phase (Section~\ref{sec:protocol}). Critic agents review the specification and tests. A human selection step locks the toy system, and the orchestrator confirms tests remain unchanged while the code passes. The structured records from this stage include the runnable pipeline, run records, and the numerical results plotted in the main text.

The \textbf{prototype experiments} node supports additional runs beyond the minimal model, for example on hardware backends. In a full campaign the orchestrator would dispatch experiment agents against the build output and reference material. For this workshop demonstration that node was not executed. The reported study ends with the validated toy pipeline and the baseline comparison.

Across the graph, each completed node leaves structured records that serve as a specification for connected descendants. Later nodes may use those records but should not silently replace them. The researcher approves steps that change scientific direction. That is the tightly controlled information flow claimed for QAOS, in contrast to a single agentic session.

\subsection{Baseline prompt}\label{sec:appB}
Baselines do not use this workflow graph and have no orchestrator, specialist, or critic chain inside a node. They receive one prompt that states the physical target and deliverables in a single message (Section~\ref{sec:baselines}). The prompt used for all five baseline runs is shown below.
\begin{quote}\small
\noindent\textbf{project description:}
\begin{itemize}\setlength{\itemsep}{0.15em}\setlength{\parskip}{0pt}
    \item Build a reproducible DFT/NEB -> localized orbitals -> embedded active-space Hamiltonian -> quantum solver -> corrected barrier workflow. - Establish an error budget for the migration barrier: geometry/functional, finite-size, embedding, basis/active-space truncation, solver and quantum-estimation error.
    \item Benchmark the quantum subproblem against the strongest tractable classical solvers (e.g., exact diagonalization / full configuration interaction (FCI) for small spaces; density matrix renormalization group (DMRG), selected configuration interaction (selected-CI), or coupled-cluster (CC) variants where appropriate).
    \item Derive hardware-aware resource estimates at a physically meaningful differential-energy target (order 10-20 millielectronvolts (meV) for room-temperature activated kinetics, subject to the material and observable).
    \item Identify a defensible crossover point: what system/active-space regime becomes classically difficult while remaining structurally favorable for quantum simulation?
\end{itemize}

\noindent\textbf{What I want:} check the input folder for relevant paper.

\medskip
\noindent\textbf{Problem scope:} proton diffusion between two fixed H2O molecules (d=2.9 angstrom), Skip embedding, active spaces, NEB and periodicity entirely: one molecular system, full electronic space, so exact diagonalization is the referee. End-to-end pipeline with DFT (structure relaxation) -> effective Hamiltonian -> solver (ED, PBE, RHF, CCSD, SQD) -> barrier correction" they should get the same files,

\medskip
\noindent\textbf{Primary deliverable:} Produce a potential-energy profile along the proton-transfer path (double-well scan):
\begin{itemize}\setlength{\itemsep}{0.15em}\setlength{\parskip}{0pt}
    \item One plot with all methods on the same axes: PBE, Hartree--Fock, CCSD, SQD, and exact (FCI) as referee.
    \item x-axis: reaction coordinate along the hop (proton position on the path).
    \item y-axis: energy (report relative to each method's minimum or a common reference, and state which).
    \item Legend: include barrier height (meV) per method.
    \item This is the main demo figure; secondary tables/figures optional.
\end{itemize}
\end{quote}

\subsection{Quantum-chemistry setting}\label{sec:appC}

The molecular model used in the demonstration is the Zundel cation, $\mathrm{H_5O_2^+}$, in which an excess proton is shared between two water molecules. Zundel-like structures are among the characteristic limiting motifs of the hydrated excess proton and occur as intermediate local structures in proton transfer through hydrogen-bonded water~\cite{schroder2022hydratedproton}. The Zundel ion is a minimal model of the actual mechanism of proton transfer in water. Nevertheless, it isolates the chemically central motif: redistribution of a proton between two neighboring oxygen centers, accompanied by a reorganization of the electronic structure.

We fix the oxygen--oxygen separation to $d=2.9$~\AA{} and define the proton-transfer coordinate $\delta$ along the line connecting the two oxygen nuclei. For each value of $\delta$, the transferring proton is constrained to this axis, while the remaining atomic coordinates are obtained from the DFT geometry optimization. The calculation thus produces a one-dimensional potential-energy profile $E(\delta)$ over a prescribed family of molecular geometries.

Proton sharing depends strongly on the donor--acceptor distance: changing the O--O separation changes the balance between proton localization on either water molecule and the symmetric, shared-proton configuration, and therefore changes the transfer barrier. Similarly, allowing the proton to move away from the O--O axis would define a different reaction path. The present scan should consequently be understood as a controlled model of proton transfer, not as a fully relaxed minimum-energy pathway.

\begin{figure}[H]
\centering
\includegraphics[width=0.75 \linewidth]{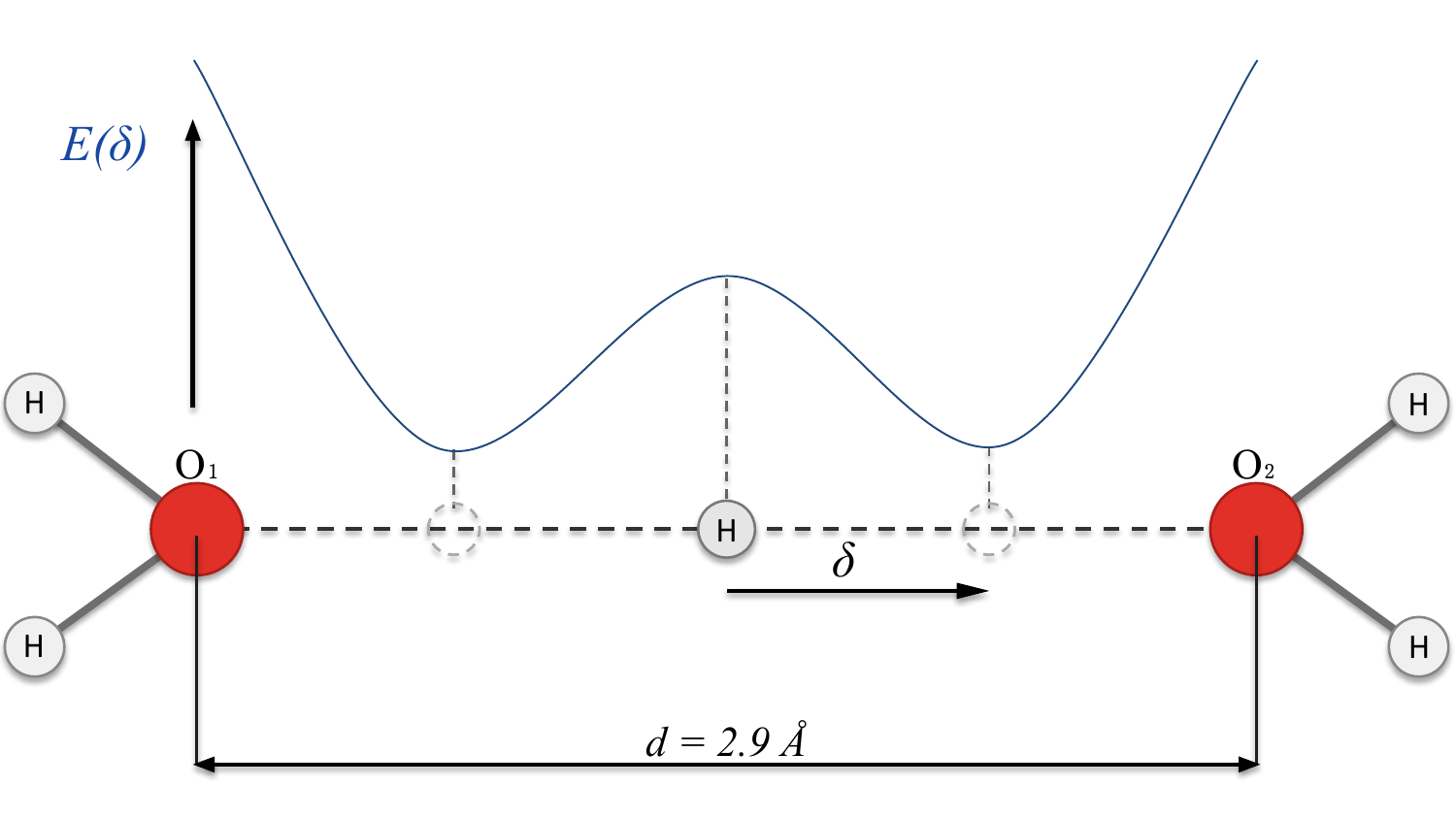}
\centering
\caption{Geometry used for the proton-transfer scan. The oxygen--oxygen separation is fixed to $2.9$~\AA{}, and the shared proton is displaced along the O--O axis according to the reaction coordinate $\delta$. The symmetric configuration lies near the center of the scan, while displacement toward either oxygen produces the two proton-localized sides of the potential-energy profile. We overlay a schematic of the potential energy as a function of $\delta$, not to scale.}
\label{fig:zundel-geometry}
\end{figure}

Within the Born--Oppenheimer description used here, the nuclei are fixed at each value of $\delta$ and the electrons are solved for that geometry. The corresponding potential energy can be written schematically as
\begin{equation}
E_{\mathrm{tot}}(\delta)
=
E_{\mathrm{nuc}}(\delta)
+
E_{\mathrm{el}}(\delta),
\end{equation}
where $E_{\mathrm{nuc}}$ is the classical Coulomb repulsion between the nuclei and $E_{\mathrm{el}}$ is the ground-state energy of the electronic terms of the molecular Hamiltonian. Both contributions vary as the proton moves, and the transfer barrier is their combined change,
\begin{equation}
\Delta E_{\mathrm{barrier}}
=
\Delta E_{\mathrm{nuc}}
+
\Delta E_{\mathrm{el}}.
\end{equation}
The nuclear term is fixed once the geometry is specified; the electronic-structure problem is to determine $E_{\mathrm{el}}(\delta)$ accurately. Consequently, when two electronic solvers are compared on exactly the same geometries and orbital basis, their difference in the predicted barrier originates from their different treatment of the electronic ground state, in particular electron correlation.

At each geometry, the nuclear positions determine the one- and two-electron integrals of the Hamiltonian in Eq.~\eqref{eq:hamiltonian}. We represent the electrons in the minimal STO-3G basis, which gives 15 spatial molecular orbitals occupied by 20 electrons. The basis itself is an approximation to the complete electronic problem, but the resulting Hilbert space is sufficiently small that full configuration interaction (FCI) remains tractable. FCI therefore supplies the exact ground-state energy of this \emph{finite-basis Hamiltonian} at each value of $\delta$, providing a controlled referee for the approximate solvers.

For larger molecules, a further reduction is normally required. Quantum-chemistry calculations often define an \emph{active space}, in which only a selected subset of electrons and orbitals is treated explicitly by the correlated solver, while chemically less relevant orbitals are frozen or otherwise removed from the many-body problem. Choosing such a space can itself be a difficult modelling problem: an active space that is adequate at one molecular geometry need not describe another geometry equally well, particularly when bonds are being formed or broken.

We deliberately remove this additional choice from the demonstration. All 20 electrons and all 15 STO-3G spatial orbitals are retained, with no frozen-core or active-space approximation. FCI, CCSD, Hartree--Fock, and SQD can therefore be compared as different treatments of the same specified molecular problem rather than of differently truncated electronic spaces. This also makes deviations in the baseline runs easy to diagnose: changing the prescribed geometry or freezing orbitals changes the Hamiltonian whose barrier is being calculated, even when the resulting computation is otherwise valid.

\paragraph{Code availability.}
We submit separate supplementary material with (i) data and plotting scripts to reproduce Figure~\ref{fig:potential-curves}, and (ii) the QAOS-built and baseline Claude-built pipelines, for direct comparison of the code each system produced from the same specification.

% The NeurIPS paper checklist is required for conference submission but is
% intentionally omitted from this arXiv preprint.

\end{document}